\pdfoutput=1

\documentclass{aastex63}

\newcommand{\bmath}[1]{\mbox{\boldmath $#1$}}

\shorttitle{Pickup ion density structure variations around the heliopause}
\shortauthors{Tsubouchi}

\begin{document}

\title{Variations in the pickup ion density structure in response to the growth of the  Kelvin--Helmholtz instability along the heliopause}

\author[0000-0003-1963-1112]{Ken Tsubouchi}
\affiliation{University of Electro-Communications, 1-5-1 Chofugaoka, Chofu, 182-8585 Tokyo, Japan}



\begin{abstract}
Features of the response of pickup ions (PUIs) to the Kelvin--Helmholtz instability (KHI) on the heliopause (HP) are examined by means of two-dimensional hybrid simulations. We assume the supersonic neutral solar wind as the source of PUIs gyrating about the magnetic field in the outer heliosheath. These PUIs become energetic neutral atoms (ENAs) via charge exchange with interstellar hydrogen, and a portion of these ENAs are detected by spacecraft such as the \textit{Interstellar Boundary Explorer} (\textit{IBEX}). To evaluate the possibility of identifying the KHI on HP from ENA observations, we assume that an imprint of the KHI may be displayed in spatial and temporal variations in the observed ENA profile. As an alternative to ENA, the column density of PUIs integrated across the HP is calculated. The KH-inducing vortex forces not only background protons but also PUIs to roll up deep in the inner heliosheath. The KH vortex also results in the emission of magnetosonic pulses that sweep PUIs in the outer heliosheath and lead to their local confinement. These effects elongate the PUIs spatial distribution in the direction normal to the HP. The appearance of the local confined structure in the PUIs column density is consequently confirmed, and this feature can be confirmed as the KHI evolution. Although the simulation cannot be quantitatively compared with the observations currently available because its resolution is too small, we expect that the derived properties will be useful for diagnosing the nature of HP fluctuation in future missions.

\end{abstract}




\section{Introduction} \label{sec:intro}
The heliosphere is formed in space as a bubble-like structure filled with solar wind plasmas (SW). Its boundary, the heliopause (HP), separates those plasmas from the local interstellar medium (LISM). The detailed physical processes in an interaction between these plasmas are of general importance and should thoroughly be investigated for understanding the termination environments of stellar winds. Recently, we had unique opportunities to observe the basic property regarding the region around the HP as a result of in situ observations by \textit{Voyager-1, 2} and energetic neutral atom (ENA) measurements with the \textit{Interstellar Boundary Explorer (IBEX)} spacecraft.

\textit{Voyager-1} (\textit{V1}) crossed the HP in August 2012 \citep{webber12, burlaga13, krimigis13, stone13} and is currently traveling through the interstellar space (\textit{Voyager-2} (\textit{V2}) in November 2018 \citep{burlaga19, gurnett19, krimigis19, richardson19, stone19}). Before completely entering interstellar space, \textit{V1} experienced a series of sudden increases and decreases of the magnetic field magnitude, coincident variations in low-energy particle fluxes, and anti-correlated variations in galactic cosmic-ray fluxes \citep{webber12, burlaga13, krimigis13, stone13}. These fluctuating results indicate that the HP is an oscillating or multilayered structure, at least in the trajectory of \textit{V1}. \citet{borovikov14} showed in their simulation results that the HP is not a smooth tangential discontinuity but is always subject to Kelvin--Helmholtz and/or Rayleigh--Taylor instabilities (KHI, RTI). \citet{swisdak13} suggested that the HP is a porous layered structure because of magnetic reconnection, which is driven by the arrival of heliospheric current sheets.

\textit{IBEX} measures the ENA flux in the energy range between 0.01 and 6 keV. Background ions are the dominant source of ENAs, which are neutralized via charge exchange with interstellar neutral atoms. Those neutral atoms are correspondingly ionized and become pickup ions (PUIs), which gyrate about the magnetic field with a velocity equivalent to the background flow speed. The velocity distribution of ENAs reflects the plasma environment of their generated region. The regime of the SW--LISM interaction can be divided into four distinct regions according to the associated plasma properties; cold and fast SW, hot and slow SW, LISM outside the HP, and unperturbed LISM. \citet{herik16} examined their three-dimensional fluid-kinetic simulations of the SW--LISM interaction and specified the ENA velocity distributions in each region.

One of the prominent discoveries by \textit{IBEX} is now known as \textit{IBEX} Ribbon \citep{mccomas09}, which is a narrow band of the enhanced ENA emission. The Ribbon is most pronounced at energy $\sim$ 1.1 keV, suggesting that the dominant source is the supersonic SW \citep[e.g., ][]{herik16}. After its discovery, several mechanisms were proposed to explain its observed properties \citep[e.g., ][]{mccomas10, mccomas14a}. It has been shown that the location of the Ribbon emission is positively associated with the local interstellar magnetic field (LISMF) $\bmath{B}$ draped on the HP, and the direction of which is perpendicular to the \textit{IBEX} line-of-sight (LOS) vector $\bmath{r}$, $\bmath{B}\cdot\bmath{r}\sim 0$ \citep{fuselier09, funsten09, schwadron09}. This association supports one of the reasonable models to account for the dominant source of the Ribbon, secondary ENA models \citep[e.g., ][]{herik10, chalov10, zirnstein15, fuselier18}. In these models, ENAs are firstly created in both the high-speed and low-temperature SW inside the termination shock and the low-speed and high-temperature SW between the termination shock and the HP (inner heliosheath, IHS). \citet{herik16} evaluated the velocity distributions of these ``primary" ENAs, where they referred to the former as neutral solar wind (NSW) and the latter as inner heliosheath neutrals (IHSN). The primary ENAs propagate outward across the HP and become PUIs in the outer heliosheath (OHS), starting to gyrate about the LISMF. When this location corresponds to $\bmath{B}\cdot\bmath{r}\sim 0$, the pitch angle of these PUIs, $\theta_{PI}$, is approximately $\sim 90^{\circ}$. If the pitch angle scattering is suppressed until the next charge exchange with interstellar neutral hydrogen atoms (timescale of a few years), these PUIs are reneutralized as secondary ENAs. By using the initial velocity distribution of PUIs originated from NSW and IHSN \citep{herik16}, \citet{roy19} examined the stability of this distribution and showed that the NSW-PUI could hold such property. Part of the secondary ENAs move straight along the LOS direction and are directly detected by \textit{IBEX} and characterized as the Ribbon. These models have succeeded in reproducing the Ribbon via global fluid-plasma/kinetic-neutral simulations, which are well fit to the \textit{IBEX} observation.

The following observations further validate the secondary ENA model. ENA fluxes detected by \textit{IBEX} normally contain those coming from the heliosheath, heliopause, and beyond, all integrated along the LOS. Therefore, it is difficult to distinguish their source regions from the data and separately analyze each property. In contrast, \citet{swaczyna16} estimated the distance to the Ribbon at $140^{+84}_{-38}$ AU from the geometrical parallax effect, indicating that the Ribbon is mainly generated just outside the HP or OHS. \citet{mccomas12} showed that emission intensities in the Ribbon ENA are ordered by both energy and latitude, with higher-energy emissions observed dominantly at high ecliptic latitudes, and low-energy emissions strongly at low latitudes. This property reflects the latitudinal structure of the solar wind around solar minimum, with the fast wind at high latitudes and slow wind at low latitudes. Therefore, the solar wind inside the termination shock (and the subsequent NSW) can be the origin of the Ribbon. 

Because the Ribbon ENA can reflect the plasma environment around the HP, dynamic processes on the HP should show an imprint on its spatial and temporal profiles. Our motivation in this study is to establish these links and to verify the possibility of using the Ribbon ENA as a probe for estimating the HP fluctuation. The Ribbon exhibits variations on timescales of about half a year, corresponding to those in the SW \citep{mccomas10, mccomas12, mccomas14b, mccomas17}. In contrast, we focused on the fluctuation properties intrinsic to the HP environment. Therefore, we should remove variations in the Ribbon associated with the SW condition. For this purpose, we assumed constant NSWs and examined the dynamics of the resultant PUIs in the OHS, which have a single gyro-velocity and can be a proxy for the observed ENAs.

Plasma flows are terminated at and deflected along the HP. Thus, there should be a velocity shear as well as a large density gradient between the IHS and OHS. Such a situation allows the growth of KHI and RTI \citep{fahr86}. At the HP nose region, the variations in the SW dynamic pressure or the momentum exchange between ion and neutral can act as effective gravity for the RTI \citep[e.g.,][]{florinski05,borovikov08,ruderman15}. The KHI appears at some distance away from the HP nose, especially when the magnetic field is perpendicular to the flow. In this configuration, the velocity shear less than twice the magnetosonic speed allows the KHI \citep[e.g.,][]{miura82}. The tangential component of the plasma velocity in the IHS, estimated by the Voyager measurement, is several tens of $\mathrm{km\cdot s}^{-1}$ on average \citep{krimigis19}. The corresponding magnetosonic speed is estimated to be $\sim 70\ \mathrm{km\cdot s}^{-1}$ from $N\sim 0.002\ \mathrm{[cm^{-3}]}$, $B\sim 0.1\ \mathrm{[nT]}$, and, $\beta\sim 1.0$ \citep{burlaga19, krimigis19, richardson19}. It indicates that the KHI can indeed take place along the HP. Its macroscopic natures have extensively been investigated in the framework of magnetohydrodynamics \citep[e.g.,][]{baranov92,ruderman93,ruderman10,borovikov14}. In this study, we paid attention to the latter situation, the KHI away from the HP nose. We aimed to examine how we can associate the KHI of the ion kinetic scale there with LOS-integrated ENA observations. By performing numerical simulations to track the individual motion of PUIs in the localized HP, we investigated the kinetic response of NSW-PUIs in the OHS to the instabilities and its manifestation in the integrated density across the HP.

Section 2 describes the simulation model and the initial setting. In Section 3, we first verify if the presence of PUIs affects the instability. Two cases having opposite density ratios between the IHS and OHS (one is realistic, while the other is unrealistic) are examined under the same pressure profiles. Next, we analyze the other case applying a larger density ratio to adapt to the more realistic HP environment. We draw the column-density map of PUIs to imitate the observation and show its relevance to the instability characteristics. Section 4 summarizes the results, discusses the limitations of this work, and proposes future research to address the limitations. We note that the present results are limited to a qualitative sense and do not aim to compare them directly with the observed properties (scales are too small to be detected).

\section{Simulation model} \label{sec:sim}
We used a two-dimensional hybrid code in the present simulations where ions are treated as individual particles and their positions and velocities are considered by solving the equations of motion. Electrons were assumed to be massless, charge-neutralizing fluids, and the magnetic fields were determined by the induction equation. The electric fields were evaluated by the generalized Ohm's law,
\begin{equation}
\bmath{E}=-\bmath{u}_i\times\bmath{B}+\frac{\nabla p_e}{eN}+\frac{\bmath{J}\times\bmath{B}}{eN},
\label{eqn:1}
\end{equation}
where $\bmath{u}_i$, $p_e$, and $N$ represent ion bulk velocity, electron pressure, and ion number density, respectively. The current density $\bmath{J}$ was given by $\nabla\times\bmath{B}/\mu_0$. In the present study, we assumed cold electrons ($T_e = 0$), so the $\nabla p_e$ term was ignored. Further detailed formulation, including the treatment of PUIs, is described in \citet{tsubo14}.

The quantities were normalized by those in the IHS (the subscripts \textit{I} and \textit{O} indicate the IHS and OHS quantities, respectively): the magnetic field, $B_I$; the proton number density, $N_I$; the Alfv\'en velocity, $v_A = B_I/\sqrt{\mu_0 m_p N_I}$; and the dynamic pressure, $m_p N_I {v_A}^2$, where $m_p$ represents the proton mass. The units of time and space were the reciprocal proton cyclotron frequency, ${\Omega_p}^{-1} = m_p/eB_0$, and proton inertial length $l_p = v_A/\Omega_p$, respectively. The spatial grid size and time step are $\Delta x=\Delta y = 0.2\ l_p$ and $\Delta t = 0.005\ {\Omega_p}^{-1}$, respectively. The two-dimensional simulation domain had $x$ and $y$ components, both having periodic boundaries, and its size was $x_{\mbox{\scriptsize{max}}}\times y_{\mbox{\scriptsize{max}}} = 102.4\ l_p \times 819.2\ l_p$, i.e., $512\times 4096$ numerical cells. To associate with the typical IHS quantities such as $N_{I}\sim 0.002$ [cm$^{-3}$] and $B_{I}\sim 0.1$ [nT] \citep{richardson08}, we correspondingly have $v_{A}\sim 50\ \text{km}\cdot\text{s}^{-1}$, $l_p\sim 5000\ \text{km}$, $\Omega_p \sim 0.01\ \text{s}^{-1}$.

The proton bulk velocity $\bmath{u}$ had the $x$-component only ($u_y = u_z = 0$). There was a velocity shear layer across the HP, with its profile given as follows,
\begin{equation}
u_x(y) = \frac{\Delta u_{x0}}{2}\tanh\left(\frac{y-y_{\mbox{\tiny HP}}}{\Delta}\right),
\label{eqn:2}
\end{equation}
where $\Delta u_{x0}$ and $\Delta$ represent the velocity difference between the velocity in the IHS and in the OHS and the half-thickness of the layer, respectively; and they were fixed to $\Delta u_{x0}=v_A$ and $\Delta = 0.8\ l_p$, respectively. The location of the HP is represented by $y_{\mbox{\tiny HP}}$. Note that we applied double periodic boundaries to this simulation model. Therefore, it is necessary to place two HP transitions in the simulation box, and we gave $y_{\mbox{\tiny HP}}=200\ (y_1)$ and $600\ l_p\ (y_2)$. The IHS corresponds to $0\le y\le y_1$ and $y_2\le y \le 819.2\ l_p$, and the OHS corresponds to $y_1\le y\le y_2$. In this setting, it is unphysical if any fluctuations from one HP are transferred into the other HP. To avoid such a situation, we made field fluctuations absorbed in the vicinity of $y_a = 0,\ 400,\ \mbox{and}\ 819.2\ l_p$ in the form of the following,
\begin{equation}
f(y) = \alpha (y) f(y)+(1-\alpha(y))f_i,
\label{eqn:3}
\end{equation}
where $\alpha(y) = -{(y-y_a)}^2/{y_w}^2 + 1$ and $f_i$ represents the corresponding initial (asymptotic) value. We applied this procedure in the region $|y-y_a|\le y_w=10\ l_p$. Hereafter, we focus on the results around the HP at $y_2$ only.

The proton density and pressure profiles were assigned by the same hyperbolic tangent form (\ref{eqn:2}) to connect the asymptotic IHS and OHS quantities as that of $u_x(y)$. The magnetic field was perpendicular to the simulation plane, i.e., $\bmath{B}=B_z\bmath{e}_z$. This property ($\bmath{B}\perp\bmath{u}$) enables the efficient growth of the KHI. The magnitude profile in $y$ was given to satisfy the static pressure equilibrium including the contribution from PUI, such as
\begin{equation}
p(y) + p_{\text{PUI}}(y) + \frac{B_z(y)^2}{2\mu_0}=\text{constant,}
\label{eqn:4}
\end{equation}
where $p(y)$ and $p_{\text{PUI}}(y)$ represent the proton and PUI pressure. The normalized proton pressure in the IHS $p_I$ is $0.5\beta_i$, where $\beta_i=p_{I}/({B_I}^2/2\mu_0)$ is the initial proton beta in the IHS. We assumed the pressure in the OHS $p_{O}=0.2\ p_{I}$ to assure the condition $B_O > B_I$ under the pressure equilibrium across the HP (\ref{eqn:4}). In simulation runs, we investigated five cases with a different combination of the OHS proton density ($N_{O}$), the PUI density ($N_{\text{PUI}}$), and $\beta_i$ (Table~\ref{tab1}).

\begin{table}[ht]
\centering
\caption{Summary of simulation runs performed in this study. $N_O$, $N_I$, and $N_{\text{PUI}}$ represent the initial number density of background protons in the OHS and IHS, and that of PUI (only present in the OHS), respectively. $\beta_i$ represents the \textbf{initial} plasma beta for the background proton in the IHS. }
\begin{tabular}{ccll}
\hline
Case & $N_O/N_I$ & $N_{\text{PUI}}/N_O$ & $\beta_i$\\
\hline
1A & 5.0 & $4\cdot 10^{-5}$ & 0.1\\
1B & 5.0 & 0.0 & 0.1\\
1C & 10.0 & $1\cdot 10^{-4}$ & 5.0\\
\hline
2A & 0.2 & $1\cdot 10^{-3}$ & 0.1\\
2B & 0.2 & 0.0 & 0.1\\
\hline
\label{tab1}
\end{tabular}
\end{table}
 
The source particle species of PUIs was assumed to be NSW-hydrogen. We should evaluate the PUI generation by calculating the charge-exchange process between protons and neutrals. However, for simplicity, we did not solve this process. Thus the number of PUIs is invariant throughout the simulation runs. All PUIs were initially assigned in the OHS, which is regarded as the source region of the secondary ENA. Therefore, in the IHS, $p_{\text{PUI}}(y)$ is explicitly zero, and $p(y)$ virtually includes the contribution of PUIs in (\ref{eqn:4}). The velocity of each PUI $\bmath{v}_{\text{PUI}}$ was given on a ring form in ($v_x,\ v_y$) space with a radius $v_r$ (the magnetic field is defined as $\bmath{B}=B(y)\ \bmath{e}_z$), such that $\bmath{v}_{\text{PUI}}=u_x\bmath{e}_x+v_r(\cos\theta\ \bmath{e}_x+\sin\theta\ \bmath{e}_y)$, where $\theta$ is randomly given and there is no thermal spread. The parameter $v_r$ represents the NSW velocity (primary ENA), the origin of PUI before its charge exchange in the OHS. We used $v_r = 10\ v_A$ in this study. Because the PUI velocity in the plasma frame is $v_r$ and its distribution is a simple ring shape, we can evaluate the normalized PUI pressure $p_{\text{PUI}}$ as $N_{\text{PUI}}\times{v_r}^2\sim N_{\text{PUI}}\times O(10)$. Though $N_{\text{PUI}}/N_O$ in Table~\ref{tab1} (especially for Case 1) is an order of magnitude lower than those used in \citet{roy19}, we gave these values to satisfy the pressure equilibrium condition (\ref{eqn:4}). 

In Table~\ref{tab1}, the simulation runs Case 1(A, B, C) represent the normal HP situation, where $N_O>N_I$. To verify the fundamental physical property in this system comprehensively, we also applied the opposite density ratio ($N_O<N_I$; no temperature gradient across the HP) in Case 2(A, B). All simulation runs were terminated when the overall system evolution was almost saturated (at $\Omega_p t_E=700$; approximately 20 hours in the present regime).

\section{Results} \label{sec:results}
\begin{figure}
\plotone{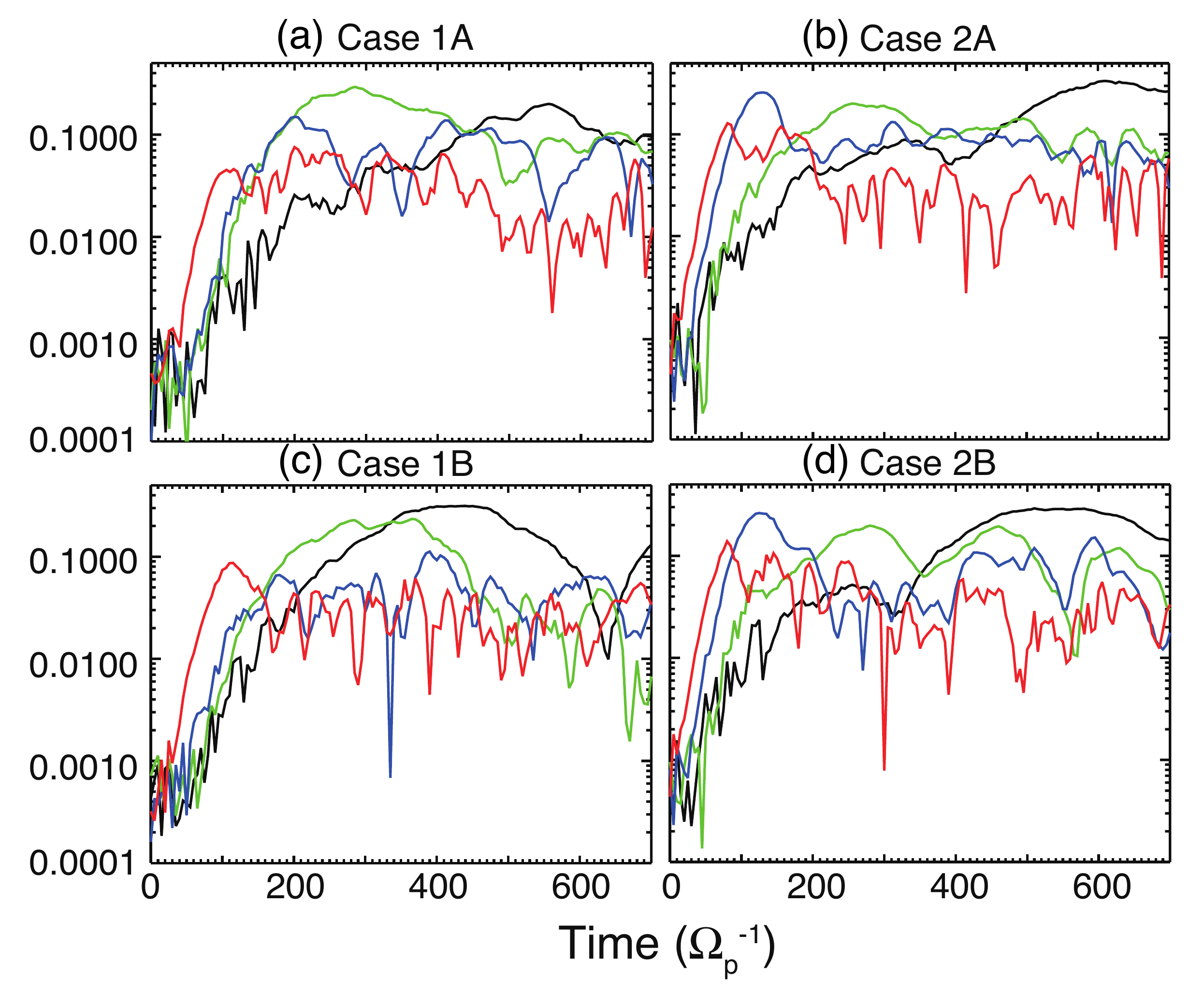}
\caption{Time variations of the Fourier mode for $\delta u_y = u_y - \overline{u_y}$. The black, green, blue, and red lines represent the wave mode for $k_x l_p = 0.061,\ 0.123,\ 0.245,$ and 0.491, respectively.}
\label{fig1}
\end{figure}

Figure~\ref{fig1} shows the growth of KH waves for (a) Case 1A, (b) Case 2A, (c) Case 1B, and (d) Case 2B given by the Fourier mode of $|\delta u_y| = |u_y - \overline{u_y}|$ along the $x$-axis at $y=y_2 (=600\ l_p)$ where $\overline{u_y}$ represents the average of $u_y$ at $y=y_2$. The black, green, blue, and red lines represent the wave modes of $m=1, 2, 4,$ and 8, where the wave number ${k_x}^{(m)} = 2\pi m/x_{\mbox{\scriptsize{max}}}$, respectively. In all cases, we can confirm the basic property of KHI evolution. The short-scale mode ($m=8$) grows at first in the linear stage. The size of the KH vortices becomes larger as they coalesce with each other. The dominant ${k_x}^{(m)}$ mode is consequently lowered until the scale becomes comparable to the system size ($m=1$). There is almost no difference in this property between the cases with and without the presence of PUI (Cases A and B). In contrast, the difference in $N_{O}/N_{I}$ (Cases 1 and 2) can be identified in their linear growth phase (until $\Omega_p t \sim 100$). The KH mode in Case 1 grows slightly slower than in Case 2, indicating that the response to the instability-driven fluctuations becomes delayed in the denser OHS plasmas (Case 1). Case 1C shows similar properties to those in Case 1A, and we do not give the corresponding plot here.

\begin{figure}
\plotone{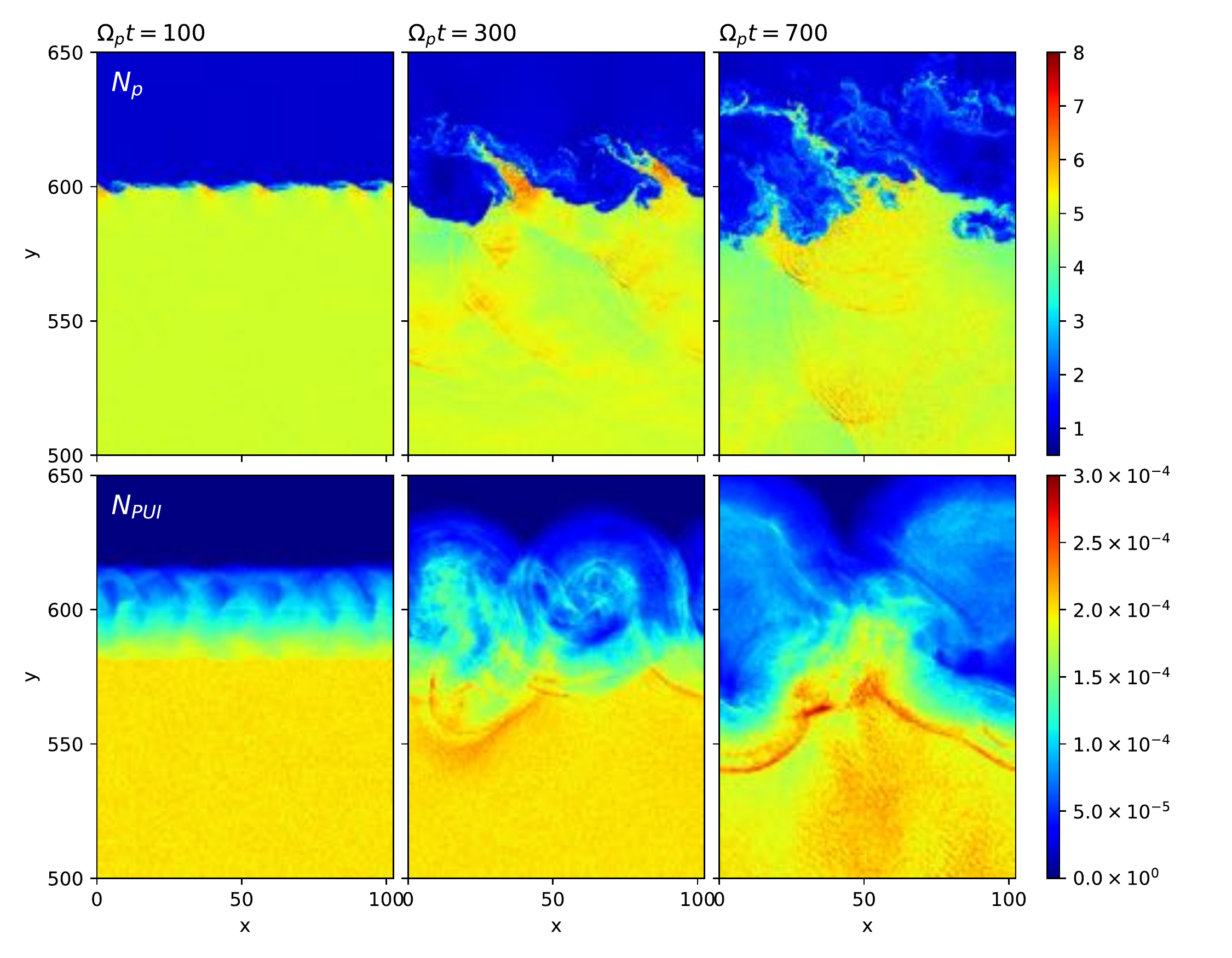}
\caption{The proton (top) and PUI (bottom) density maps in the $x-y$ plane for Case 1A. From left to right: the snapshot at $\Omega_p t=$100, 300, and 700.}
\label{fig2}
\end{figure}

\begin{figure}
\plotone{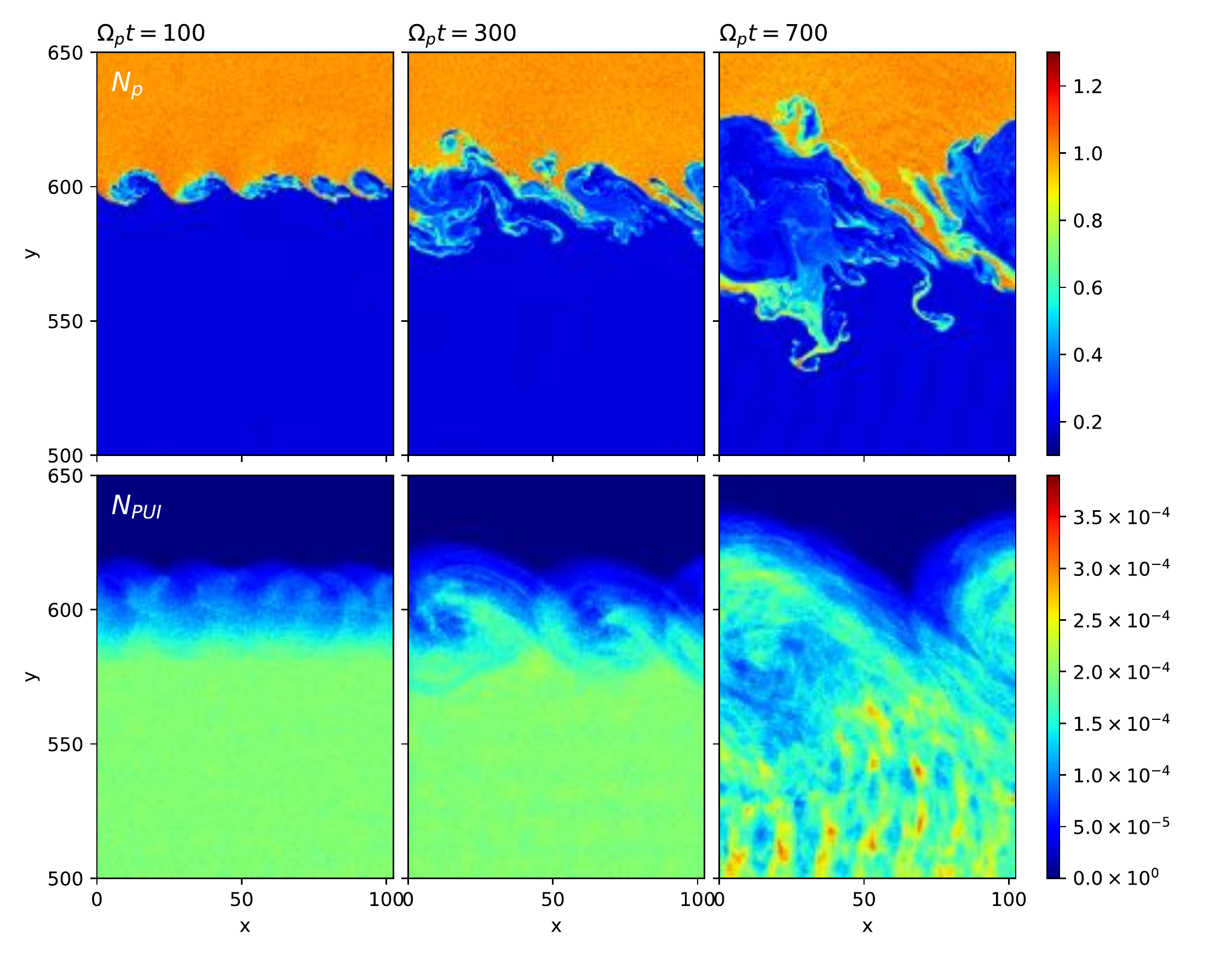}
\caption{The proton (top) and PUI (bottom) density maps in the $x-y$ plane for Case 2A. From left to right: the snapshot at $\Omega_p t=$100, 300, and 700.}
\label{fig3}
\end{figure}

We show the temporal development of the density structures for protons (top) and PUIs (bottom) in Figure~\ref{fig2} for Case 1A and Figure~\ref{fig3} for Case 2A. The snapshot time is (from left to right) $\Omega_p t=$ 100, 300, and 700. The KH properties indicated in Figure~\ref{fig1} are confirmed in the proton density ($N_p$) profiles (upper panels): the delayed growth in the linear stage for Case 1A, the transition from small- to large-scale fluctuations by the coalescence, and the appearance of the small-scale turbulence at the edge of the vortices \citep[e.g., ][]{matsumoto04, matsumoto06}. The vortices are formed in the lower density region and not in the lower thermal pressure region (in the IHS for Case 1A and the OHS for Case 2A), which is typical of RTI. In Case 1A (Figure~\ref{fig2}), local confinement close to the vortex root can also be identified at $\Omega_p t=300, 700$, whereas such a structure cannot be seen in Case 2A.

In the PUI density ($N_{\text{PUI}}$) profiles (bottom panels), both cases 1A and 2A show uniformity in the OHS, and the vortex profile in the transition region is well correlated with that of protons at $\Omega_p t=100$ (KH linear stage). For Case 1A, PUIs are more diffuse in the IHS than protons in the nonlinear stage ($\Omega_p t=300, 700$). At $\Omega_p t=700$, the KH vortex carries PUIs deep in the IHS. In addition, the localized increase in $N_{\text{PUI}}$ can be confirmed as the thread-like region along the HP and the region broadly spread in the OHS. In contrast, Case 2A indicates that the PUI spatial distribution almost coincides with that of protons and is confined in the OHS until the end of the simulation run. At $\Omega_p t=700$, periodic monotonic structures appear in the OHS, which cannot be seen in Case 1A.

\begin{figure}
\plotone{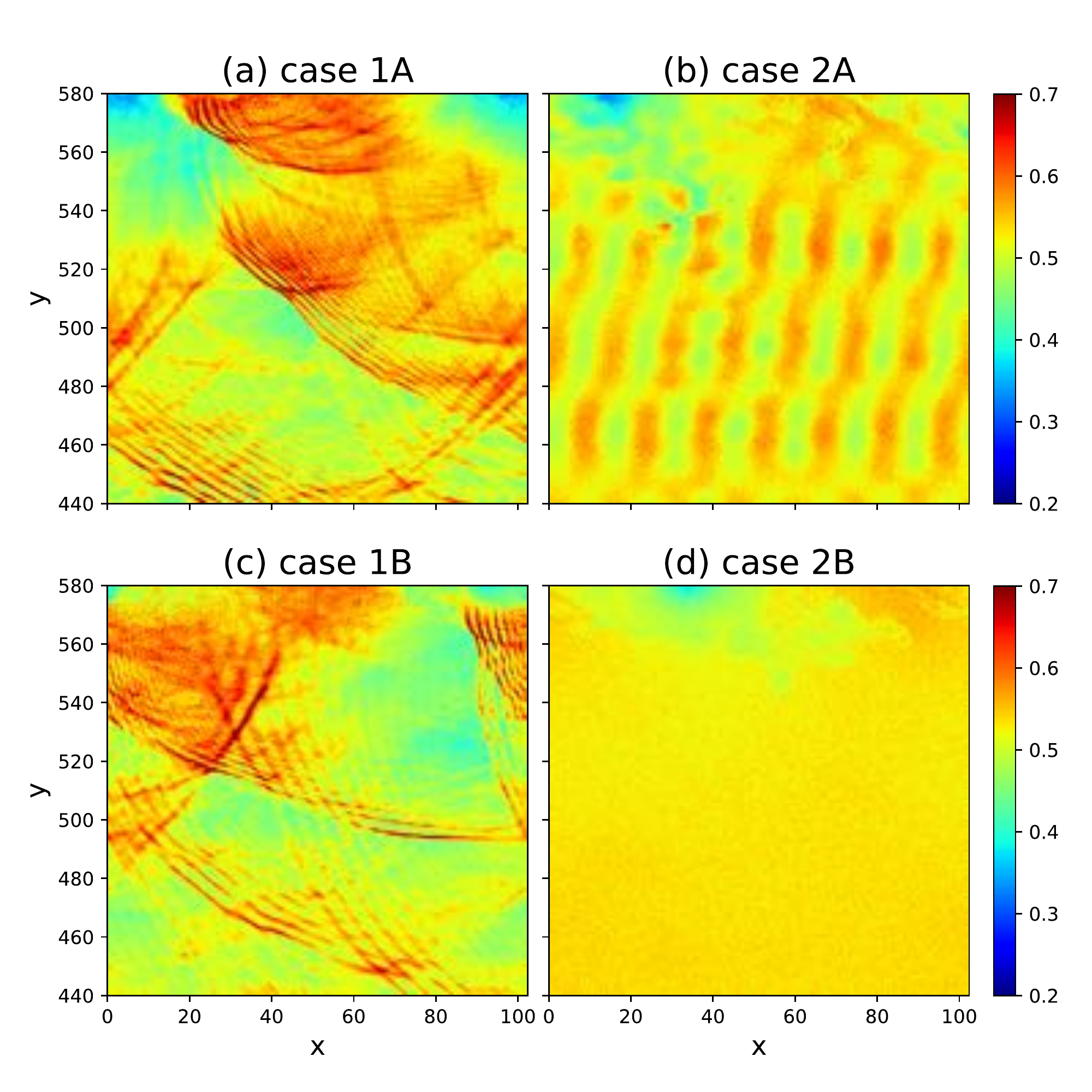}
\caption{Close-up views of the magnetic pressure in the OHS at $\Omega_p t=700$ in each run: (a) Case 1A, (b) Case 2A, (c) Case 1B, and (d) Case 2B.}
\label{fig4}
\end{figure}

Figure~\ref{fig4} shows the close-up views of the magnetic pressure in the OHS ($440\le y\le 580$) at $\Omega_p t=700$ from each run, (a) Case 1A, (b) Case 2A, (c) Case 1B, and (d) Case 2B. The condition of $N_{O}>N_{I}$ (Case 1A/B) allows the emission of a train of magnetosonic pulses in response to the KH wave evolution at the HP (Figure~\ref{fig4}a and \ref{fig4}c), which are also confirmed in the proton density profile in the top panels of Figure~\ref{fig2}. Magnetosonic waves are induced where the plasma flow in the KH vortices collides with the HP interface. Because the OHS is a high Mach number regime, these induced fluctuations are effectively amplified during their propagation and the resultant pulse emission occurs. As PUIs are swept by these pulse propagations, the local density concentration in the OHS shown in Figure~\ref{fig2} is formed. In this condition, PUI dynamics hardly affect the OHS environment. In contrast, the condition of $N_{O}<N_{I}$ (Case 2A/B) leads to different magnetic field configurations between the case with and without PUIs (Figure~\ref{fig4}b and \ref{fig4}d). While the magnetic field in the OHS is uniform in the case of no PUIs (Case 2B, Figure~\ref{fig4}d), the presence of PUIs results in the generation of monotonic waves (Case 2A, Figure~\ref{fig4}b), which are also identified in the $N_{\text{PUI}}$ map (Figure~\ref{fig3}).

\begin{figure}
\plotone{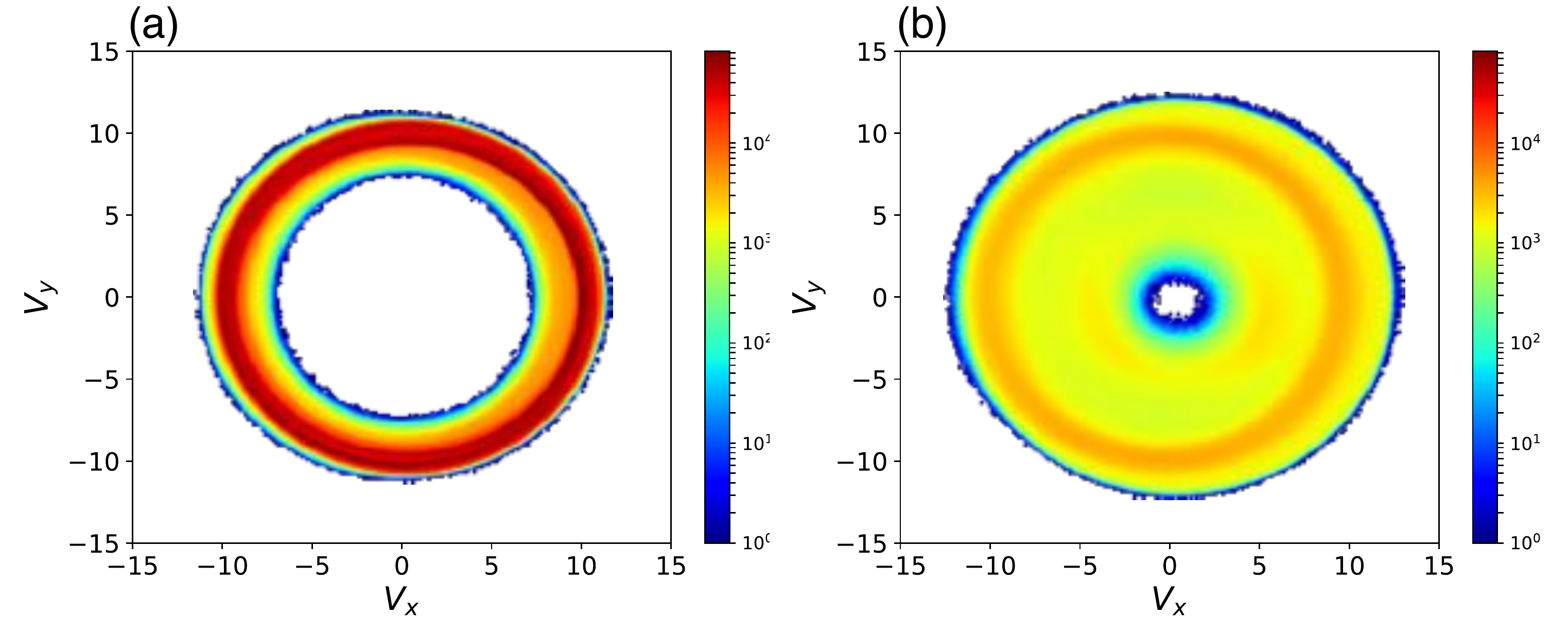}
\caption{PUI velocity distribution maps in the $(v_x, v_y)$-plane at $\Omega_p t=700$: (a) Case 1A, (b) Case 2A.}
\label{fig5}
\end{figure}

Because the total proton inertia in the OHS is weak in this low $N_{O}$ condition, the free energy in the PUI velocity distribution is large enough to induce such fluctuations. In Figure~\ref{fig5}, the distribution maps in the $(v_x, v_y)$-plane for PUIs appeared in the area of Figure~\ref{fig4}a and \ref{fig4}b are shown. While the initial ring-shaped distribution is relatively well maintained in Case 1A (Figure~\ref{fig5}a), the distribution in Case 2A (Figure~\ref{fig5}b) is strongly scattered in energy by the self-induced fluctuations shown in Figure~\ref{fig4}b. Correspondingly, it can be expected that the PUI energy is well conserved in Case 1A and becomes diffuse in Case 2A.

\begin{figure}
\plotone{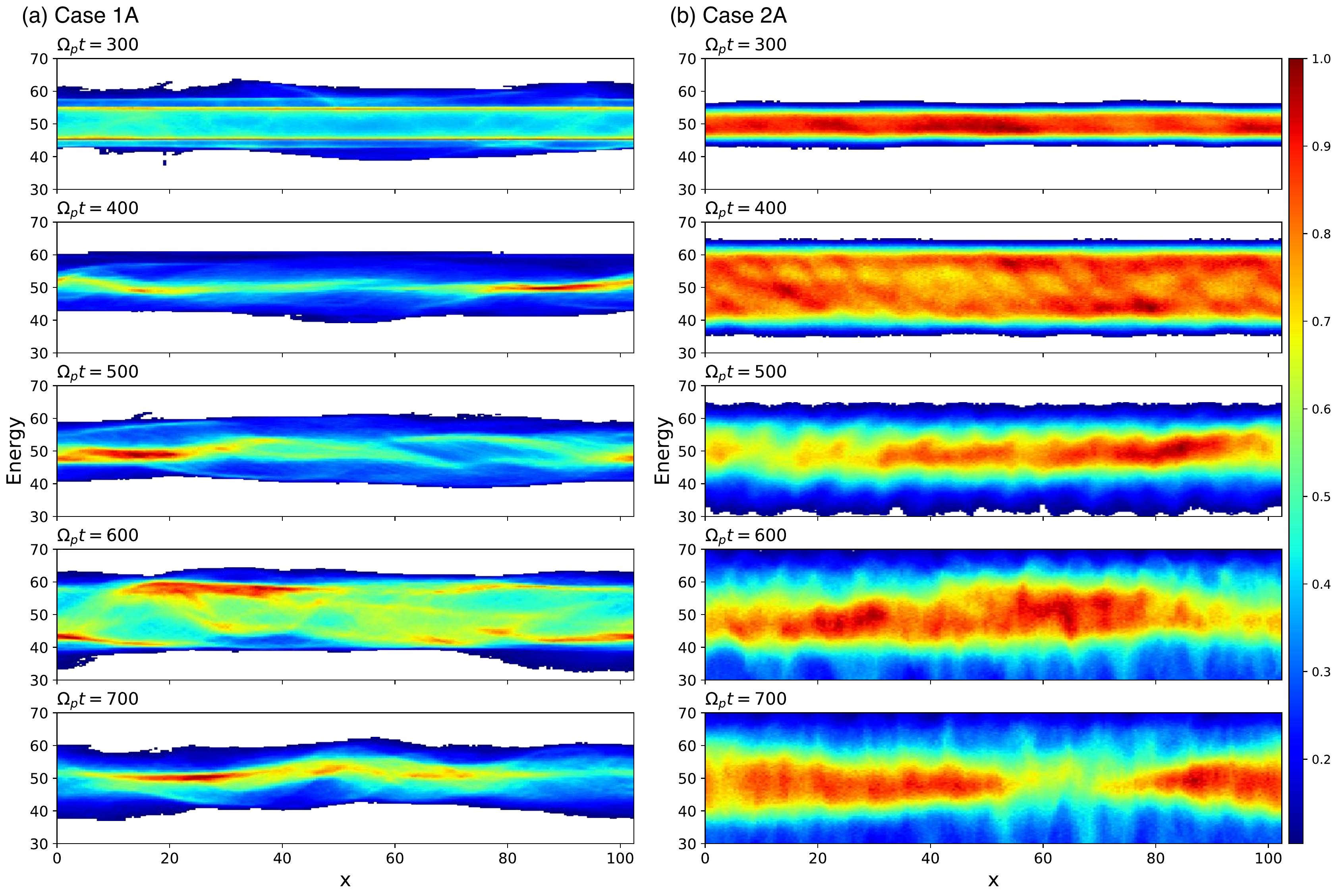}
\caption{The $x$-profiles of the PUI column density per energy integrated over the $y$-axis between $460\le y\le 700\ l_p$ for (a) Case 1A and (b) Case 2A. From the top to the bottom: $\Omega_p t = 300,\ 400,\ 500,\ 600$, and 700. The color code represents the normalized counts of PUIs in a linear scale. The counts less than 10\% of the maximum in each panel are shown blank.}
\label{fig6}
\end{figure}

\textit{IBEX} detects the ENA flux as the column density along the LOS direction. In the present simulation system, the $x$-axis is parallel to the HP surface and the $y$-axis can be regarded as the virtual LOS direction. To associate the simulation results with the observation, we made a column density data of PUI integrated over the $y$-direction between $460\le y\le 700\ l_p$. In Figure~\ref{fig6}, the PUI density per energy divided by 200 bins between $30\le E\le 70\ m_p {v_A}^2$ is shown for (a) Case 1A and (b) Case 2A.

The density profile is almost uniform both in space and energy at $\Omega_p t = 300$ for Case 1A (top panel of Figure~\ref{fig6}a), where KH-induced fluctuations are weak and have little effect on the PUI dynamics. The width of the energy band ($45\sim 55$) simply corresponds to the PUI velocity gyrating about the magnetic field ($v_r \pm 0.5 \Delta u_{x0}$). This energy spread remains largely unchanged throughout the simulation run, suggesting that the PUI energy diffusion is weak in Case 1A. From $\Omega_p t =$ 300 to 700 (except 600), the local density enhancement near $E_l\sim m_p {v_r}^2/2=50\ m_p {v_A}^2$ and its spatial transition is confirmed. This feature is because of the PUI accumulation in the OHS as well as the distribution of PUIs elongated into the IHS, both of which are indicated in Figure~\ref{fig2} and can be associated with the consequence of KH wave evolution. For Case 2A (Figure~\ref{fig6}b), in contrast, the energy range in the column density is broadened and the structure becomes uniform as time proceeds. This diffusion, both in energy and space, is because of the velocity scattering, as shown in Figure~\ref{fig5}. Consequently, we can no more recognize the footprint of KH from this column density profile.

\begin{figure}
\plotone{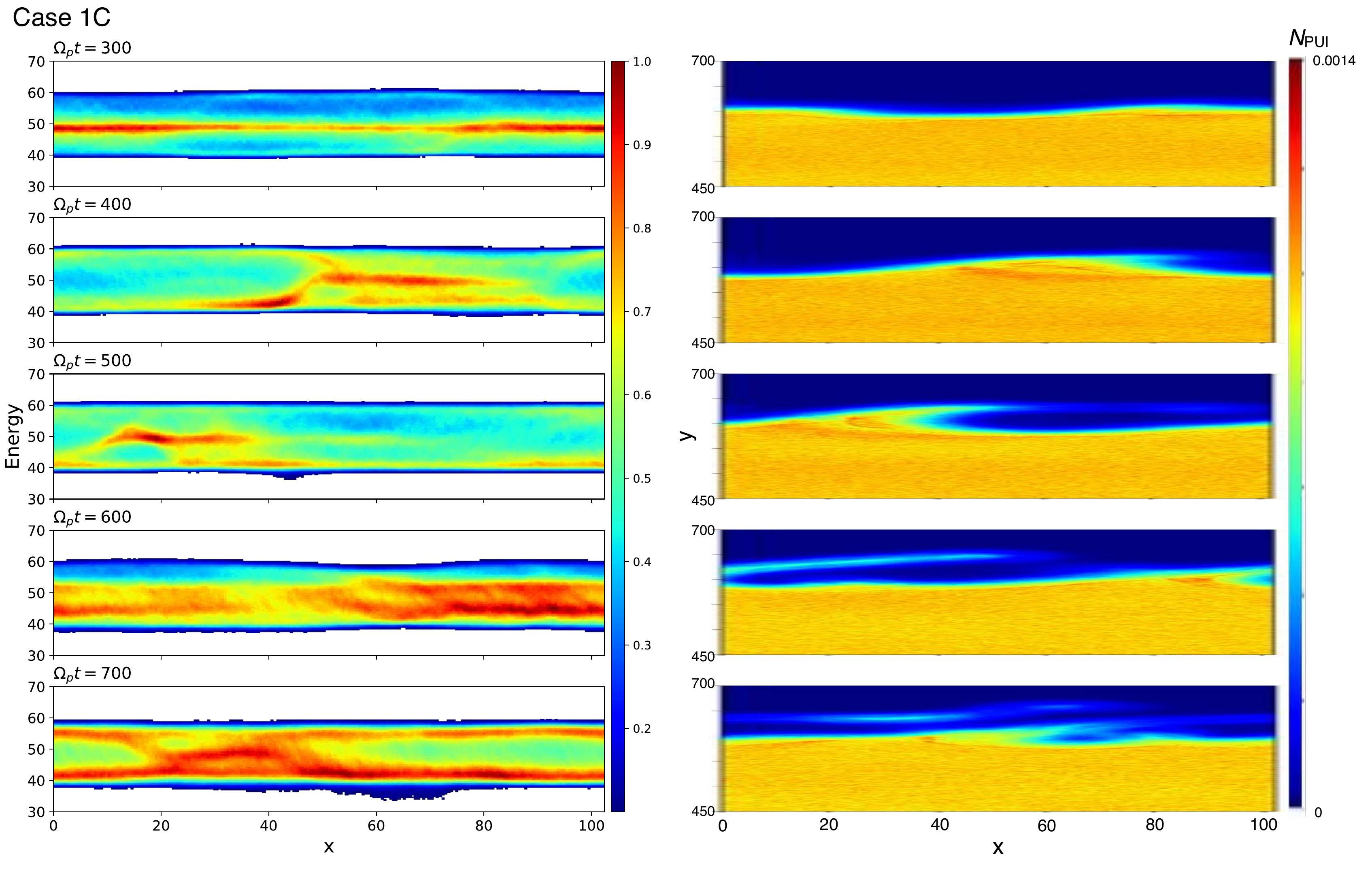}
\caption{Time evolution profiles from $\Omega_p t=300$ to 700 for Case 1C: (left) the PUI column density (the same format as Figure~\ref{fig6}), (right) the PUI density maps in the $x-y$ plane. }
\label{fig7}
\end{figure}

In Case 1A and 2A, we verified the fundamental KHI property in the presence of PUI, especially its dependence on the density ratio $N_{O}/N_{I}$. We gave this ratio about an order of magnitude lower than real to save the computational cost. $\beta_i$ was also lower than real because we did not take PUIs in the IHS into account. To adapt these parameters to the realistic HP as possible, we use $(N_{O}/N_{I},\ \beta_i)=(10.0,\ 5.0)$ in the run Case 1C. Figure~\ref{fig7} shows the results: (left) the spatial profiles of the PUI column density from $\Omega_p t=$ (top) 300 to (bottom) 700 in the same format as Figure~\ref{fig6}, (right) the corresponding PUI density maps in the $x-y$ plane.

Similar to Case 1A (Figure~\ref{fig5}a), the PUI velocity distribution retains a ring shape with weak diffusion (not shown here). We can confirm this feature in the left panels of Figure~\ref{fig7}, where the energy spread ($40\le E\le 60\ m_p{v_A}^2$) remains unchanged throughout the run. In contrast to Case 1A (Figure~\ref{fig4}a), the plasma beta in the OHS is higher ($\sim 0.25$), so the KH vortex-induced fluctuations subsonically propagate. Thus, we can not identify any wave steepening and the following magnetic pulse emission in the OHS (not shown here). Correspondingly, there is no local confinement of the PUI density in the OHS away from the HP (the right panels of Figure~\ref{fig7}).

In the column density maps (the left panels of Figure~\ref{fig7}), we can confirm the localized enhancement near $E_l\ (\sim 50\ m_p{v_A}^2)$ and its displacement along the $x$-direction. These features are common to both Case 1A and 1C. As mentioned above, such a density enhancement mainly came from the following two processes in Case 1A. One is the PUI accumulation swept by the magnetic pulses in the OHS. The other is the elongation of the OHS in the LOS direction induced by the KH vortex evolution. In Case 1C, the former process is almost ineffective. Therefore, we expect that Case 1C can link the localized column density enhancement to the KHI more directly than Case 1A.

The upper right panel ($\Omega_p t=300$) of Figure~\ref{fig7} shows the HP slightly inflates into the IHS in the range $0\le x\le 20$ and $70\le x\le 100$, but its fluctuation amplitude is small. In these regions, the corresponding column density (the left panel) is relatively higher than in the surroundings. However, the overall spatial profile roughly looks uniform. After $\Omega_p t=400$, the KH vortex rolls PUIs deeply toward the IHS as it moves in the transverse $(-x)$ direction. The propagation speed is approximately $0.4\ v_A$. The localized increase in the column density at around $E\sim E_l$ (the left panels) can be indirect evidence of the KHI occurrence because it is identified in the region inflating toward the IHS. The energy $E_l$ is almost equivalent to the PUI velocity $\bmath{v}_{\text{PUI}}\sim \pm v_r\bmath{e}_y$, implying that the PUI motions are dominantly along the LOS direction. It further indicates that the KHI evolution induces this motion. 

\section{Summary and Discussion} \label{sec:summary}
In this study, we performed two-dimensional hybrid simulations to investigate the PUI density profiles in the vicinity of the HP. There must be a velocity shear and a density jump across the HP and thus KHI and/or RTI grows. The main aim of this study is to explore how such instabilities on the HP can be identified by ENA observations, such as those detected by \textit{IBEX}. We specifically focused on the KHI along the HP away from its nose region. \citet{mccomas14a} referred to the KHI as one of the source mechanisms of generating the Ribbon ENA because the confinement of plasmas in KHI-generated structures could produce enhanced ENA fluxes. However, they also pointed out that this scenario does not explain the geometry of the Ribbon. Accordingly, we did not pursue associating the KHI with the mechanism of the ENA generation further. Rather, our interest was that variations in the Ribbon ENA profile will display an imprint of the instability as long as the secondary ENA model is finally validated and the detected ENA is mainly generated in the OHS. For this purpose, we investigated the characteristics of the PUI distribution associated with the HP instabilities in the situation before the ENA generation. The difference in the plasma density ratios between OHS and IHS, $N_{O}/N_{I}$, for normal ($>1$) and abnormal ($<1$) cases was verified. The results are summarized as follows:
\begin{enumerate}
\item There is little difference in the KHI property whether PUIs are present in the OHS. However, it should be noted here that we ignored both the charge-exchange process and PUIs in the IHS. These must substantially affect the instability. Further evaluation by taking these effects into account will be necessary.
\item In the normal HP situation, the KH wave forces PUIs to roll up deep in the IHS in accordance with its growth. It also emits intermittent magnetosonic pulses in the OHS, which sweep PUIs and lead to localized accumulation. Consequently, the PUI spatial distribution is elongated in the direction normal to the HP, resulting in the local enhancement of the PUI column density integrated along the virtual LOS.
\item In the abnormal HP situation, monotonic magnetosonic waves are induced in the OHS, leading to the diffusion of PUI velocities in the component perpendicular to the magnetic field. The column density, in this case, shows a resultant energy broadening, and its total profile becomes almost spatially uniform.
\item In the realistic situation (higher $N_{O}/N_{I}$ and $\beta_i$), the local enhancement of the PUI column density is more directly related to the KHI location. Thus, we expect that this feature represents a sign of the KHI occurrence on the HP.
\end{enumerate}

Case 1A (the normal HP) shows weak diffusion in the PUI energy and the displacement of its strong emission in the $x$-direction, i.e., along the HP. This feature is emulated in the ENA map that would be observed and can be used as evidence of the growth of the KH mode. In contrast, Case 2A suggests that the KH mode in the unrealistic density ratio ($N_{O}/N_{I}<1$) cannot be identified in the map. As shown in these cases, the column-density map is strongly sensitive to the background plasma density (or temperature) ratio between the IHS and OHS, even in the same pressure profile. In addition, the density ratio in Case 1A, $N_{O}/N_{I}=5$, might be much lower than that found in the Voyager observations, where $N_{O}$ is estimated to be several tens of $N_{I}$ \citep[e.g.][]{krimigis13, gurnett13}. To examine the case of this ratio (and also the plasma beta) higher than Case 1A, we showed tentative results in Case 1C and confirmed basic features commonly seen. Much larger computational resources will be necessary to verify the case using the realistic density ratio. It is an issue for future work.

The parameter sets used in this study were only specific cases, and it is necessary to examine other conditions, especially for constructing the initial PUI velocity distribution, more extensively to understand the general properties of PUI dynamics associated with the KHI/RTI over the HP. At least in the present condition, the PUI dynamics hardly affect the HP global features in response to the instability. This result is because of the dominance of the magnetic pressure around the HP environment compared with the thermal one. We cannot further estimate the accurate contribution of PUI to the pressure balance. In the case of the PUI pressure dominating the magnetic one, the KH property itself may be modified. We will leave this evaluation for future investigation.

Another simplification is that PUIs were only present in the OHS and their initial velocity distribution was represented by a ring shape about the magnetic field with no thermal spread ($\bmath{v}_{\text{PUI}}$ in the $x-y$ plane and $|\bmath{v}_{\text{PUI}}|=v_r=10\ v_{A}$). This ring distribution is generally considered to be unstable and generates waves, which scatter the PUI pitch angle to form a shell shape. In contrast, for the ENA emission restricted to the Ribbon form, the initial ring distribution in the OHS should be stably maintained in a timescale of charge exchange for secondary PUIs, typically over years. Because parallel fluctuations (the $z$-component) were suppressed in the present field configuration where the magnetic field was perpendicular to the simulation plane, we cannot evaluate the effect of parallel diffusion and the consequent pitch angle scattering for PUIs. The validity of our results might be restricted to the condition that the PUI velocity distribution of a ring shape is stably kept in the simulation as long as possible.

Recently, \citet{florinski16} performed theoretical and numerical studies to investigate PUI-ring stability. They showed that the scattering process was far more complex than previously thought and that the ring distribution retained its shape in a specific regime. They also showed that magnetic fluctuations can trap PUIs in a regime of pitch angle $\sim 90^\circ$ and lead to spatial confinement that provides sufficient ENA flux for the Ribbon \citep[e.g.,][]{schwadron13, isenberg14, giacalone15, zirnstein19}. The mechanism of the spatial confinement of PUI itself is still the fundamental and significant issue to understand the Ribbon properties. A three-dimensional simulation will be necessary in the future to examine how parallel diffusion alters the column density profile of PUIs associated with KHI and RTI, as indicated in this study.

In the PUI column density, the contribution from the IHS should also be considered, but this was ignored in the present study. \citet{fuselier18} showed the Ribbon identified at 0.2 keV, indicating that its source is the shocked solar wind in the IHS. As \citet{herik10} showed in their numerical computations using the magnetohydrodynamic-plasma/kinetic-neutral model, using a Lorentzian (or kappa) distribution is more accurate for IHS protons, where PUIs correspond to a power-law tail part, representing suprathermal components accelerated at the termination shock. The next step is to include this population explicitly in the simulation model and to verify the energy dependence on the consequent profile in accordance with the growth of the KH mode.

Spatial and temporal variations in the \textit{IBEX} Ribbon must mainly be followed by those in the solar wind. However, our simulation results suggest that instabilities over the HP also induce such variations manifested in the PUI column-density map, even when the solar wind condition is kept constant. By applying typical IHS quantities in the present parameters, the time and spatial scales of these features correspond to the order of tens of hours and thousands of kilometers ($\sim 0.01$ au). In contrast, the source region of the line-integrated Ribbon flux extends a few tens of au from the HP \citep{zirnstein15}. Therefore, these resolutions are obviously too small to be detected by \textit{IBEX} instruments. The results shown in this study cannot directly be associated with any observed features available at present. However, because of the limited simulation domain, the evolution of the KH vortex is saturated in these simulations. The KHI must continue to progress further and can result in the formation of a much larger vortex \citep[e.g.,][]{borovikov14}, which might be accessible in the observation. Thus, we would not deny the applicability of the present results to the observed phenomena.

The global map of ENA flux is possibly the only source to display the macroscopic HP properties. By verifying the characteristics of the PUI column density for a number of parameter sets (not only KH-associated but any probable plasma configuration), we can expect to develop a template to identify the physical processes occurring on the HP from the observed data in future missions.

\acknowledgments
The computer simulation was performed using the KDK computer system at the Research Institute for Sustainable Humanosphere, Kyoto University, and the computational resources provided by the Institute for Space-Earth Environmental Research (ISEE), Nagoya University. The computation was also performed on the FX100 supercomputer system at the Information Technology Center, Nagoya University. This work is supported by JSPS KAKENHI Grant number JP17K05666.

\end{document}